\documentclass[twocolumn,prb,aps]{revtex4}
\usepackage{epsfig}
\begin{document}
\setcounter{totalnumber}{10}

\title{Simulations of binary hard-sphere crystal-melt interfaces: 
interface between a one-component fcc crystal and a binary fluid mixture}


\author{Rachel Sibug-Aga and Brian B. Laird\cite{brian} \\
        Department of Chemistry, University of Kansas, 
        Lawrence, Kansas 66045, USA}

\date{\today}

\begin{abstract}
The crystal-melt interfaces of a binary hard-sphere fluid
mixture in coexistence with a single-component hard-sphere crystal 
is investigated using molecular-dynamics simulation.  In the
system under study, the fluid phase consists of a two-component 
mixture of hard spheres of differing size, with a size ratio $\alpha=0.414$. At low pressures
this fluid coexists with a pure fcc crystal of the larger particles in which the small
particles are immiscible. For two interfacial orientations, [100] and [111], 
the structure and dynamics within the interfacial region is studied and compared with
previous simulations on single component hard-sphere interfaces. 
Among a variety of novel properties, it is observed that
as the interface is traversed from fluid to crystal the diffusion
constant of the larger particle vanishes before that of the small
particle defining a region of the interface where the large particles
are frozen in their crystal lattice, but the small particles exhibit
significant mobility.  This behavior was not seen in previous binary hard-sphere
interface simulations with less asymmetric diameters. 
\end{abstract}
\maketitle

\section{Introduction}

A fundamental understanding of the nucleation, growth kinetics and morphology 
of crystals grown from the melt requires a detailed microscopic description of
the crystal-melt interface\cite{Woodruff73,Tiller91,Howe97,Adamson97}. However,
such interfaces are very difficult to probe experimentally and 
reliable experimental data, especially for structure and transport properties, is 
rare. It is then not surprising that computer simulations have, in recent years, 
played a leading role in the determination of the microscopic structure,
dynamics and thermodynamics of such systems\cite{Laird98}. 

To date, the vast majority of simulation studies have focused on single 
component interfacial systems. Such studies range from simple model systems such as hard
spheres\cite{Kyrlidis95,Mori95,Laird98,Davidchack00} or Lennard-Jones\cite{Broughton86c,Galejs89} 
to more ``realistic'' systems, such as water\cite{Karim88,Karim90,Hayward01}, 
silicon\cite{Abraham86,Landman86} or simple metals\cite{Jesson01,Hoyt01}. 
In contrast, there have been but few studies on multicomponent
systems\cite{Davidchack96,Davidchack99}, in spite of the fact that most materials of technological 
interest are mixtures (for example, doped semiconductors, alloys and intermetallic compounds). 
In such systems, the crystal and coexisting fluid have differing composition, in general, and
the change in concentration as one traverses the interface from one bulk phase into the other 
becomes an object of study. 

Of particular interest to materials scientists
is the degree of interfacial segregation - the preferential adsorption of one component 
(usually the ``solute'') at the interface.  In addition, the phase diagrams for multi-component
systems are significantly more varied and complex than single component systems due to the 
additional dimension of concentration. For a binary system several types of solid-liquid 
equilibria are possible.  If the two types of particles are similar, then one typically has
coexistence between a binary fluid and a substitutionaly disordered solid of similar 
structure to that of the pure components. However, if the two types of particles are substantially
different in nature, then generally the binary fluid will either be immiscible in the pure
coexisting solid, or will coexist with one or more ordered crystal mixtures (e.g. intermetallic
compounds).  Previous simulation studies on binary crystal-melt interfaces have exclusively focused on 
the former case, namely the equilibrium between the fluid and a disordered crystal.  Davidchack and 
Laird\cite{Davidchack99} recently reported results for a binary hard sphere system in which a substitutionally 
disordered face-centered-cubic (fcc) crystal coexists with a binary fluid mixture. 
In a related study, Hoyt {\it et al.} examined the crystal-melt interface of a 
Cu/Ni mixture\cite{Hoyt01}. In both studies the degree of solute segregation was found to be negligible. 

In the two studies above, the disordered fcc crystal was stabilized by the fact that the two components
were quite similar in size -for example, in the hard-sphere system studied by Davidchack and Laird, the
diameters of the two types of spheres making up the system differed only by 10\%. 
In this work, we extend the previous studies to hard-sphere mixtures with significant size
asymmetry. For such systems, in which the diameters differ by more than about 85\%, the disordered
fcc phase is no longer stable and only coexistence of the fluid with ordered crystal structures 
is possible.  In this work  we examine the interface between a binary hard-sphere fluid mixture and a
coexisting fcc crystal in which the small particle is immiscible. 

Our system of choice is a binary hard-sphere mixture in which the ratio of the smaller particle 
diameter to that of the larger particle is 0.414. Hard spheres are an important reference 
system for the crystal-melt interfaces of simple systems since the structure, dynamics and phase behavior
of dense atomic systems are dominated by packing considerations with only minor influence from the attractive
parts of the interactions. For example, it has been recently demonstrated\cite{Laird01} that the interfacial
free energy of close-packed metals can be quantitatively described using a purely hard sphere model.
The specific diameter ratio of 0.414 was chosen because, to 
perform an interface simulation, accurate phase coexistence parameters are required {\it a priori}, and the
phase diagram for this binary system has been worked out via simulation in some detail\cite{Trizac97}. This
phase diagram shows that at low pressures the fluid mixture coexists with a pure fcc crystal of
the larger particles, but that at higher pressures the crystal structure in equilibrium with
the fluid is a 1:1  (or AB) ``intermetallic'' compound with an ``NaCl'' structure (the small and large particles
form interpenetrating fcc lattices). (The existence of the ``NaCl'' structure at this diameter ratio had  been
predicted earlier, using cell theory\cite{Cottin95}.  The diameter ratio, $\alpha=0.414$ is necessary 
for an ``NaCl'' structure to attain its maximum packing fraction of 0.793.) Thus, this system allows us to
study the interfaces between binary fluids and two types of ordered crystal phases: single component
and ``NaCl''. In this work we present results for the former, but simulations on the 
fluid/``NaCl'' are under way and will be reported later. 
\section{Description of the System}
We consider a two-component system consisting of hard spheres
of differing diameters, given by $\sigma_A$ and $\sigma_B$. Without loss of
generality, it is assumed that $\sigma_A \ge \sigma_B$. 
The interaction between two particles of type $i$ and $j$, ($i,j \in \{A,B\}$), 
respectively, is then given by 
\begin{equation}
\phi_{ij}(r)=\left\{\begin{array} 
             {r@{\quad,\quad}l}          
             \infty & r\le\sigma_{ij}  \\ 0 & r>\sigma_{ij}
             \end{array} \right.,
\end{equation}
where $r$ is the distance between the centers of the two interacting spheres, and
$\sigma_{ij}$ is the distance of closest possible approach.  In addition, we define the spheres to be additive, that is, 
$\sigma_{ij} = (\sigma_i + \sigma_j)/2$.  The state of the system is then completely 
described by specifying the total density, $\rho = \rho_A + \rho_B = N/V$, the mole fraction,
$x_A$, of the larger species, and the diameter ratio $\alpha = \sigma_B/\sigma_A$. Note that
so defined one has $\alpha \in (0,1)$.  In a single component system composed of hard spheres of diameter $\sigma$ 
the packing fraction, $\eta$ (the fraction of the total volume occupied by the spheres) is given by,
\begin{equation}
\eta=\frac{\pi}{6} \rho \sigma^3,
\end{equation}
where $\rho$ is bulk density.  For the binary hard sphere system  described above
the packing fraction is 
\begin{eqnarray}
\eta & = & \eta_A + \eta_B \\
     &=&\frac{\pi}{6} \rho \left[x_A \sigma_A^3 + x_B\sigma_B^3\right]\\
     &=&\frac{\pi \sigma_A^3}{6} \rho [x_A  + (1-x_A)\alpha^3] \;  .
\end{eqnarray}

\begin{figure}[h!]
\epsfig{file=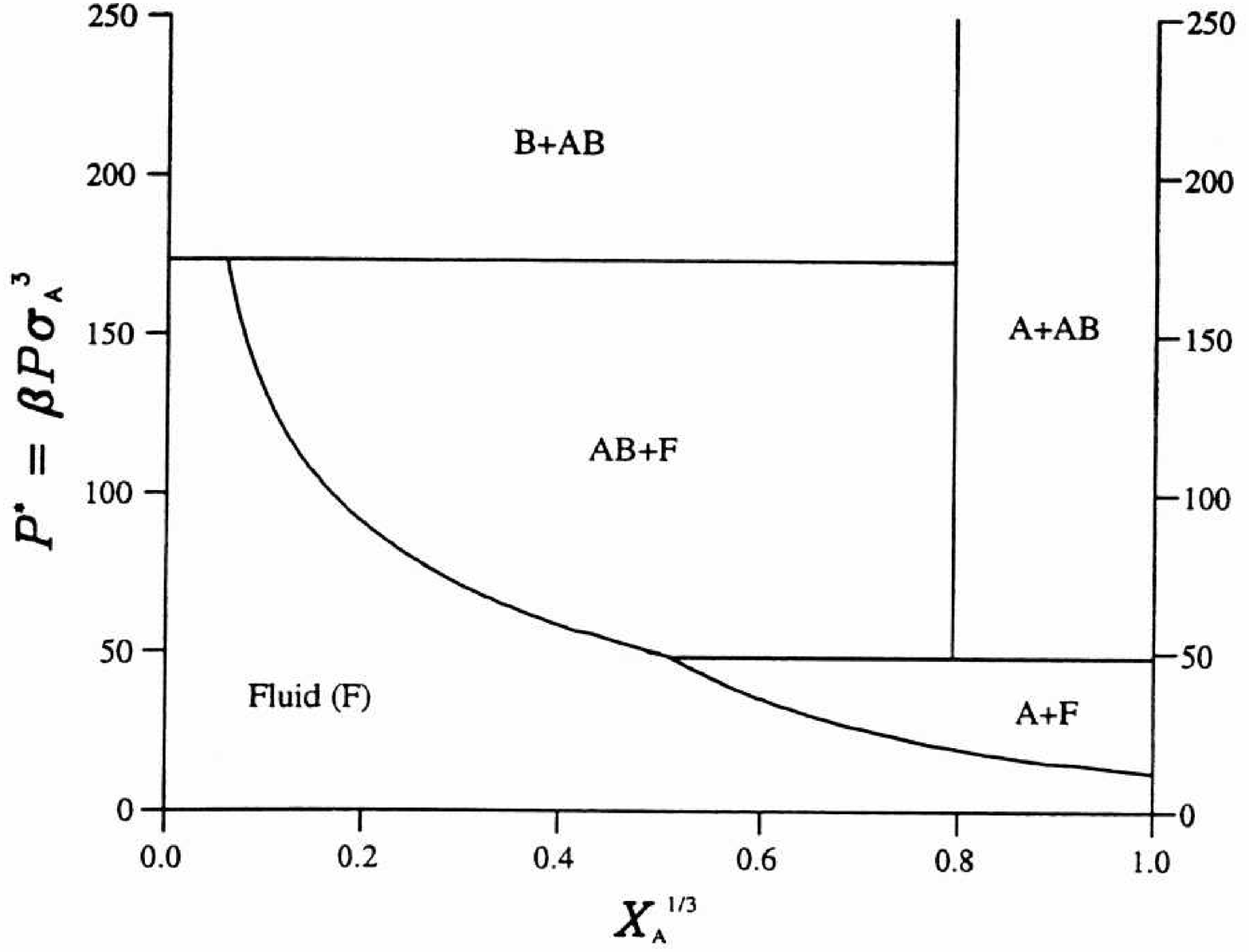,height=5.6cm,width=8.0cm,angle=0}
\caption{\small Pressure-concentration phase diagram of a binary hard-sphere
system with $\alpha=0.414$.  [Reprinted from Ref. 22, by permission 
of the publisher, Taylor and Francis Ltd. (www.tandf.co.uk/journals).]  
Note that to make the phase coexistence lines easier to distinguish, the pressure is plotted against $x_A^{1/3}$ 
and not $x_A$ as in the usual case.}
\label{phasediag}
\end{figure}

As mentioned in the introduction we are interested in the present study in 
the interface between an fcc crystal consisting of pure large (type A) spheres and its coexisting 
binary fluid at a diameter ratio, $\alpha=0.414$.  The pressure-composition phase diagram for a binary 
hard sphere system with this diameter ratio has been previously determined by Trizac and coworkers\cite{Trizac97} 
and is shown in Fig.~\ref{phasediag}.

For this study we have chosen the point in the phase diagram where a  fluid mixture with a 1:1
composition (that is,  $x_A=0.50$)  coexists with a crystal phase that is an fcc composed 
of only large particles.   We independently calculated the coexistence conditions for this 
point in the phase diagram and we determine the coexisting pressure to be 
$P = 20.1\sigma_A^3/kT$, 
with packing fractions for the crystal and fluid calculated to be $\eta_c=0.61$ and $\eta_f=0.51$, respectively.

\section{Calculation of Interfacial Profiles}

To monitor changes in the structural or dynamical properties across the 
interface, the system is divided into bins along the $z$-axis, defined
as that perpendicular to the interfacial plane.  Quantities of interest are then 
calculated for each bin generating a $z$-dependent interfacial profile for
the specific property (density, concentration, diffusion, etc.) being measured.  
The techniques of profile generation and analysis are similar to those used earlier
in the works of Davidchack and Laird on the single\cite{Davidchack98} and binary hard-sphere 
systems (with $\alpha = 0.9$)\cite{Davidchack99}. In this section these techniques are
summarized, with particular attention to the present calculation. The reader is urged to
consult the earlier papers if more detail is required.

In our analysis of the simulations, we employ bins of  two different resolutions: a coarse scale and 
a fine scale.  Coarse scale bins have a width equal to the layer 
spacing of the bulk crystal.  This spacing is $0.753 \sigma_A$ for [100] 
and $0.870 \sigma_A$ for [111].  The fine scale is $1/25$ of the coarse scale.  Fine scale 
bins reveal in greater detail parameter variations across the interface, but the coarse scale  
is more useful for observing overall trends in the interfacial profiles. Also for some
parameters, such as the diffusion constant, only the coarse scale can be used if one is to
achieve meaningful statistical accuracy. For interfacial profiles that exhibit oscillations
on the order of the lattice spacing, such as density, the conversion between the fine-scale profiles and
coarse-scale profiles to illustrate bulk trends is problematic, since the distance between the
peaks of such profiles is not necessarily constant through the interface. The mismatch 
between the coarse-scale bins and the peak spacing can lead to spurious results\cite{Davidchack98}
if one simply averages over the fine scaled bins to create the coarse scaled profile.  For such
profiles we employ a Finite Impulse Response filtering procedure\cite{NumRec} to average out
the oscillations and reveal coarse-grained trends. Details of the specific filtering procedure
we use can be found in Ref. 24.  

Below is a description on how the various interfacial properties were 
determined.  In the definitions, the size of the bin is denoted by $\Delta z$ and 
$L_x$, $L_y$ and $L_z$ are the the dimensions of the simulation box in the $x$, $y$ and $z$ 
directions, respectively.

\begin{itemize}
\item {\bf \it Pressure:} The total pressure profile is defined as
\begin{equation}
P(z)=\frac{1}{3}\left\{P_{xx}(z)+P_{yy}(z)+P_{zz}(z)\right\}\; ,
\end{equation}
where$P_{kk}$ is calculated from
\begin{equation}
\frac{P_{kk}}{k_BT}=\rho(z)+\frac{3m}{2L_xL_y\Delta z\Delta t\!\!<E_k>\!\!}
\sum_{c=1}^{N_c} r^{(c)}_k \Delta v^{(c)}_k,
\label{pkk}
\end{equation}
where $c$ indexes the collisions, $m$ is the mass of each sphere, $\!\!<E_k>\!\!$ is the average kinetic energy 
per sphere, $N_c$ is the number of collisions that occurred over the time 
interval $\Delta t$ in the region between $z-\Delta z$ and $z+\Delta z$, 
$r^{(c)}_k$ is the $kth$ component of the relative distance between the two 
colliding spheres and $\Delta v^{(c)}_k$ is the $kth$ component of the change in 
velocity for collision $c$.  The first term in Eq.~\ref{pkk} represents the 
ideal gas pressure and the 
second term is the excess part due to sphere interactions.
\item{\bf \it Excess Stress Profiles:} The local excess stress is calculated from the pressure tensor components. 
\begin{equation}
S(z)=P_{zz}(z)-\frac{1}{2}\{P_{xx}(z)+P_{yy}(z)\}
\end{equation}
In a simulation of an equilibrium interfacial system this quantity should be zero, except in a small region at
the interface. Improper preparation or equilibration of the system often manifests itself in the 
excess of this quantity in the bulk crystal away from the interface. As such this quantity is carefully monitored
as a measure of the quality of the simulation. To smooth out the large oscillations in this quantity through
the interface, the profile is filtered to easily reveal overall trends.  (The local excess stress can be integrated 
with respect to $z$ to give the surface excess stress. For a liquid-vapor interface the surface excess stress 
is identical to the interfacial free energy, but since the relaxation time for stress in a crystal is generally 
much longer than a typical simulation time, the surface excess stress and $\gamma_{sl}$ can be significantly different 
for crystal-melt interfaces\cite{Tiller91}.) 
\item {\bf \it Density Profiles and Contour Plots:} The fine-scale density profile for a sphere of type $i$ is 
determined from the number density of that type particle in each fine-scale bin.
\begin{equation}
\rho_i(z)=\frac{<\!\!N_i(z)\!\!>}{L_xL_y\Delta z}
\end{equation}
where $<\!\!N_i(z)\!\!>$ is the average number of spheres of type $i$ in the 
region between $z-\Delta z/2$ and $z+\Delta z/2$. To observe overall trends in
bulk density (or concentration) changes we also produce filtered density profiles using 
our FIR filtering procedure discussed above. 

In addition to the $z$ dependent density profiles, it is also useful to examine the density variations within
the {\it x-y} planes parallel to the interfacial plane. To do this we divide the system into orthorhombic subcells
with a width in the $z$ direction equal to the coarse bin spacing and $x$ and $y$ dimensions of 
$0.15 \sigma_A$. By counting the average number of particles of each type in each subcell and dividing
by the subcell volume we can produce 2d contour plots of the cross-sectional density variation within
each interfacial plane. 
\item{\bf \it Interface location:} We determined the location of the interface from the orientational 
order parameter profile.
\begin{equation}
q_n(z)=\left<\frac{1}{N_z}\sum_{i,j,k}cos\left\{n\theta_{xy}(i,j,k)\right\}\right>
\label{qn}
\end{equation}
where $n$ is an integer, $i,j$ and $k$ are nearest neighbor atoms, 
$\theta_{xy}(i,j,k)$ is the bond angle formed by $i,j$ and $k$ projected on 
the 
$x,y$ plane, and $N_z$ is the total number of atoms that form  bond angles. 
The average is taken over the number of angles found between $z-\Delta z/2$ and 
$z+\Delta z/2$.
The interface in the [100] orientation is the point along the $z$-axis where 
$q_4$($q_6$ for the [111]) is the arithmetic mean of the bulk crystal and liquid 
values. For comparison, the position of the Gibbs dividing surface\cite{Tiller91} is also 
calculated.  We determine the Gibbs dividing surface as the plane along the z-axis such 
that for the 'solute' $i$, $\Gamma^i=0$ in the equation
\begin{equation}
N^i/A=\rho_S^iz+\rho_L^i(L_z-z)+\Gamma^i
\label{gibbs}
\end{equation}  
where $N^i$ the total number of spheres of type $i$, $A$ is the area of the 
interface, $\rho_S^i$ and $\rho_L^i$ are the bulk densities, $z$ is the 
location of the interface assuming the length of the simulation box runs 
from $0$ to $L_z$ and $\Gamma^i$ is the excess particle per unit area of 
the interface.
\item{\bf \it Diffusion coefficient profile:} To study the dynamics across the interface, the diffusion 
coefficient profile is calculated.  For a particle of type $i$, the diffusion coefficient is 
defined as follows
\begin{equation}
D_i(z)=\frac{1}{6N_i(z)}\frac{d}{dt}\sum_{j=1}^{N_i(z)}
\left<{\mbox{\boldmath$r$}_j(t)-\mbox{\boldmath$r$}_j(t_0)}^2\right>
\end{equation}
The term in the summation is the mean squared displacement over a time 
interval 
$t-t_0$ of a total of $N_i$ type $i$ spheres located between $z-\Delta z/2$ and 
$z+\Delta z/2$ at time $t_0$.
\end{itemize}

\section{Construction and Equilibration of Interface}

Initially, blocks of crystal and fluid spheres at the calculated coexistence packing 
fractions and concentrations were prepared separately.  As a reference,
the $z$-axis is taken to be perpendicular to the interface.  
The {\it x-y} planes for both blocks had the same dimensions so that they would fit perfectly 
when put together to construct the interface.  The plane perpendicular to the interface is 
made as close to square as possible given the geometric constraints of the specific 
interfacial orientation under study.  This is trivial to achieve with the [100] orientation 
but for [111], the $x$ and $y$ lengths can only be made approximately equal.  
The lengths along $z$ were made longer than both those in $x$ and $y$ so that bulk properties 
will be observed between the two interfaces formed. Periodic boundary conditions are applied
in all directions, which results in the two independent crystal-melt interfaces 
formed along $z$. The  similarity of the two interfaces is an important monitor
on the quality of the simulation. Obviously, if statistically significant differences in 
structure or dynamics exist between the two interfaces, then the system has not been 
properly equilibrated. 

The crystal block with [100] orientation was set up with 7776 large spheres.  
It consisted of 48 crystal layers, each layer having 162 spheres.  Using the coexistence packing 
fraction $\eta_c=0.61$, the following dimensions for the [100] crystal block 
were used: $L_x=13.56\sigma_A$, $L_y=13.56\sigma_A$ and $L_z=36.15\sigma_A$.   Its coexisting fluid 
had 7776 large spheres and 7776 small spheres (15552 spheres total).  The block length is $L_z=43.78\sigma_A$.  For reasons
that will be explained later, this $L_z$ gives a packing fraction that is slightly higher than that obtained 
from the calculated coexistence conditions.  For the simulation of the [111] interface, the  crystal block used 
contained  8190 large spheres, with 45 layers in the $z$ direction giving 182 spheres per layer.  The crystal 
block dimensions are $L_x=13.85\sigma_A$, $L_y=12.91\sigma_A$ and $L_z=39.13\sigma_A$.  
The total number of fluid spheres used was also 15552 as 
in that for the [100] simulation with  $L_z=45.09\sigma_A$,  again, giving a packing fraction slightly higher than that
predicted for coexistence. Thus the total number of particles in the interface simulations are  23328 and 23742 for
the [100] and [111] interfacial orientations, respectively. 

Both crystal and fluid blocks are equilibrated separately.  The two blocks are 
then put together but a gap equal to $\sigma_A$ is left between each of the 
two crystal-melt interfaces formed to ensure that no initial overlap 
will occur at the interfaces.  The molecular dynamics simulation is then started with only the 
fluid spheres allowed to move (the crystal spheres remain fixed). The  fluid then fills  the gaps. To 
compensate for the decrease in the overall bulk density of the fluid phase during this step, the fluid blocks 
are prepared at a packing fraction that is slightly higher than the predicted coexistence values (as 
mentioned earlier).  In the next step, the crystal is equilibrated with the fluid spheres held fixed.
At this point the interface setup  is complete and an equilibration run is started with all spheres
moving and with  initial velocities assigned according to a Maxwell distribution.  In order to efficiently
carry out the molecular dynamics simulation of such a large system we use the cell method
of Rappaport~\cite{Rappaport95}.

The stability of a crystal-melt interface in a simulation is extremely sensitive to
the assumed coexistence conditions. In our previous work\cite{Davidchack98,Davidchack99}, it
was found that the predetermined coexistence conditions generally had to be modified
slightly in order to create a stationary interface with a zero excess stress in
the bulk crystal region. This is necessary because a) the coexistence conditions are often
not known {\it a priori} to the accuracy required for interface stability and b) the presence
of the interface in a finite simulation can shift the coexistence equilibrium slightly. 
During our preliminary runs for the current system, using the coexistence conditions 
as calculated by thermodynamic integration of the free energies of separate bulk phases, 
we found that the resulting interface was stable, but yielded a bulk crystal with negative excess
stress.  Through experimentation, we found that an equilibrium interface with zero crystal
excess stress was possible if the initial fluid packing fraction was increased to $\eta_f=0.52$.   
This had the effect of changing the concentration equilibrium slightly away from a 1:1 
mixture in the fluid, as discussed below. Now it is in principle possible to vary both the
initial fluid concentration and packing fraction so that the final equilibrium gives precisely
a 1:1 fluid mixture, however this procedure is quite tedious and since our choice of the 1:1
fluid at coexistence was arbitrary, the fact that the actual system deviates slightly from
this concentration is not important for the purposes of the current study. 

\begin{figure}[!h]
\epsfig{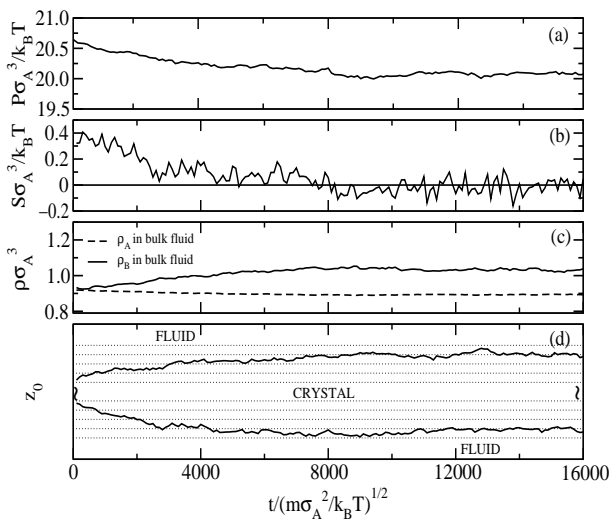}
\caption{\small Time evolution of (a) total pressure of the system,
(b) stress in bulk crystal, (c) fluid densities and (d) location of the 
interfaces, $z_0$. The time unit $(m\sigma_A^2/k_BT)^{1/2}$ 
corresponds roughly to 18 collisions per particle (cpp).}
\label{timeev}
\end{figure}

To ensure that the system is indeed in equilibrium and that the bulk crystal is free of excess stress, 
we monitor a variety of properties such as total pressure, bulk crystal stress, fluid bulk densities and 
interfacial location.  The results for the [100] interface are shown in Fig.~\ref{timeev}, which
shows that prior to equilibration at about $t^{*} \equiv  t (kT/m\sigma_A^2)^{1/2} = 10000$
the crystal grows by about 3 crystal lattice planes (see Fig.~\ref{timeev}d), 
accompanied by a pressure drop from 20.6 to its equilibrium value of 20.1 $\sigma_A^3/k_BT$ 
(Fig.~\ref{timeev}a).  In addition, the average excess stress in the bulk crystal, initially positive, goes to 
zero (within flucuations) when equilibrium is reached (Fig.~\ref{timeev}b). (This average excess stress
was calculated by averaging $S$ as defined above over the middle 28 layers of the bulk crystal.)
  
Initially, the bulk densities of both 
particle types in the fluid are equal, but as the system equilibrates, 
the bulk density of the small particles increases.  This 
increase  is due to the growth of the crystal (see Fig.~\ref{timeev}d).  Large 
fluid particles near the crystal freeze, expelling the small particles, which are  immiscible
in the crystal at this pressure, into the bulk fluid region.  
Although the bulk fluid initially has a large sphere mole fraction of $X_A=0.50$, the value at
equilibrium is somewhat lower (0.46 and 0.47  for the [100]and [111] interfaces, respectively). 
The equilibrium packing fraction of the bulk fluid slightly reduced from its initial value of
$\eta_f=0.52$  to 0.51.  Once the system is in equilibrium, the interfacial positions
are stable and the fluctuation in position is less within one layer spacing.     

In the preparation of the [100] interface some small particles became trapped within some of
the interior crystal layers as the crystal grew during equilibration. Since these were in regions
where the diffusion constant for the small (and large) particles was found to be zero, it
cannot be determined whether these particles would actually be present in a true equilibrium
interface. In order to determine the importance of these interstitial small particles in 
stabilizing the interface, we removed the particles (about 77 total) from the inner 3 
crystal layers where they were found. The removal was done at $t^* = 8000$ in the
equilibration run. Initially the crystal stress became negative, but quickly returned
to zero (within fluctuations) as small particles from the bulk diffused in to reoccupy
the removed layer closest to the interface (this layer corresponds to layer B in
Fig.~\ref{f100rho}, discussed in the next section). The inner two layers did not fill in. 
The interfacial position remained stable during this process.  The question of true
chemical equilibrium is always a tricky one in these types of interface 
simulations\cite{Davidchack99} due to the extremely slow relaxation of concentration
in the deeper crystal layers. However, in this region the concentration of
small particles is in any event probably quite small and should not affect our
results significantly (except for perhaps the interfacial segregation). 
As a possible check to this procedure, one could use the Widom insertion method~\cite{Frenkel96} to
determine the excess chemical potential, and thus the  solubility, of the small particles
in the various inner crystal layers, but this was not done here.

\begin{figure}[!h]
\epsfig{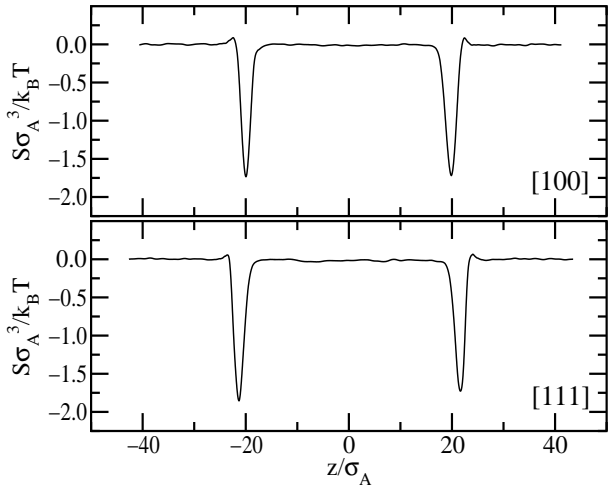}
\caption{\small Filtered excess stress profiles for the [100] and [111] interface orientations. \hfill}
\label{fstr}
\end{figure}

The total length of the averaging run after equilibration was $t^* = 4000$, 
which was divided into 40 separate blocks of length $t^*=100$ (corresponds to about 
1800 collisions per particle(cpp)),over which the interfacial profiles were averaged.  
Since the system contains two interfaces, each block average yields
two independent profiles (when properly folded about the center of the crystal)
Thus, each of the profiles reported here represents an average of 
80 block averages.

It is important to compare the two independent interfaces produced in
a single interface simulation to ensure that they are statistically 
identical. Significant differences between the two interfaces are indications
of problems with the equilibration procedure. As a diagnostic we determine the
excess stress profile (calculated on the fine scale and filtered using the FIR
filter described above and in Ref. 24). 
These filtered stress profiles are shown in Fig.~\ref{fstr} for both the 
[100] and [111] orientations - note that, the crystal is
in the middle of the simulation box.  
The profiles are remarkably symmetric and also show that the excess stress is
zero within fluctuations in the bulk crystal region.  
It should be noted that 
in contrast to the case for a liquid-vapor interface, the interfacial free energy of a 
crystal melt interface {\it cannot} be determined from the integral of the excess stress profile,
as the relaxation time for stress in the crystal is  significantly longer than possible
simulation times\cite{Tiller91}, and must be determined by other means, such as the recently developed
cleaving wall method\cite{Davidchack00}. The excess stress profiles shown 
in Fig.~\ref{fstr} show
a significant negative stress region on the crystal side of the interface, indicating that in 
this region the transverse pressure components are greater than the pressure component 
normal to the interfacial plane. The precise origin of this unrelaxed 
crystal stress at the interface is as yet unknown.

\begin{figure}[h]
\epsfig{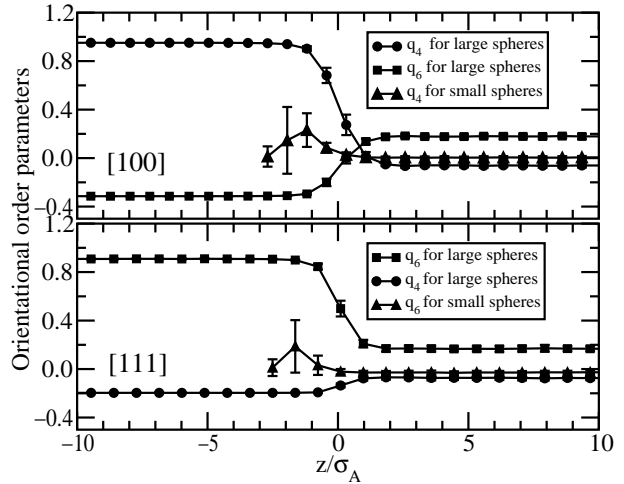}
\caption{\small Orientational order parameters  $q_4$ and $q_6$ for the two sphere 
types and for both interfaces studied. \hfill}
\label{ford}
\end{figure}

As mentioned above, the position of the interface is determined as the
value of $z$ at which the orientational order parameter for the large
spheres is the arithmetic mean of that quantity in the two bulk phases. This
quantity is a useful measure of interfacial location as it is monotonic as
a function of $z$ (so that using the arithmetic mean makes sense) and can
be calculated as smooth function without large fluctuations using relatively
short simulation runs.  Orientational order parameters $q_4$ and $q_6$, as defined 
by Eqn.~\ref{qn}, were determined for each particle type.  These are shown in 
Fig.~\ref{ford}.  Since the crystal phase is made 
up of pure large spheres and we want to see how the ordering of particles 
is changed from bulk crystal to bulk liquid, we determined the interface 
location from the parameters calculated for large spheres.  
We also show  $q_4$ and $q_6$ for the small particles and we see that at 
the interfacial region, the small spheres start developing some order 
that is similar to the large spheres.

\section{Results for the [100] and [111] Interfaces}

\subsection{Structure}
\begin{figure}[h]
\epsfig{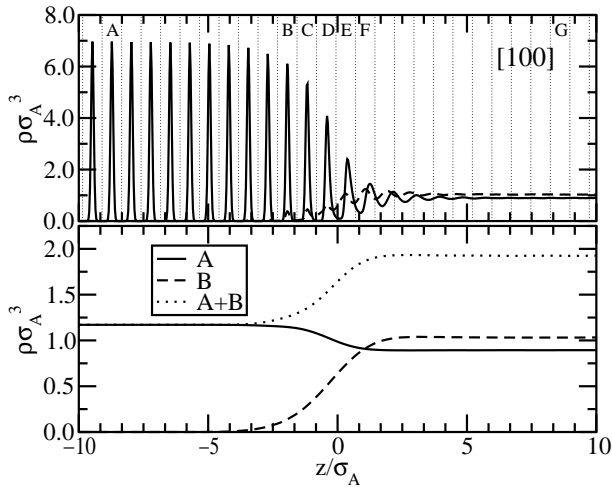}
\caption{\small Fine-scale (upper panel) and filtered (lower panel) density profiles for the 
[100] orientation. The solid line and dashed lines are for the larger (A) and smaller (B) particles,
respectively. In the lower panel the dotted line shows the filtered total 
density. \hfill}
\label{f100rho}
\end{figure}

\begin{figure}[h]
\epsfig{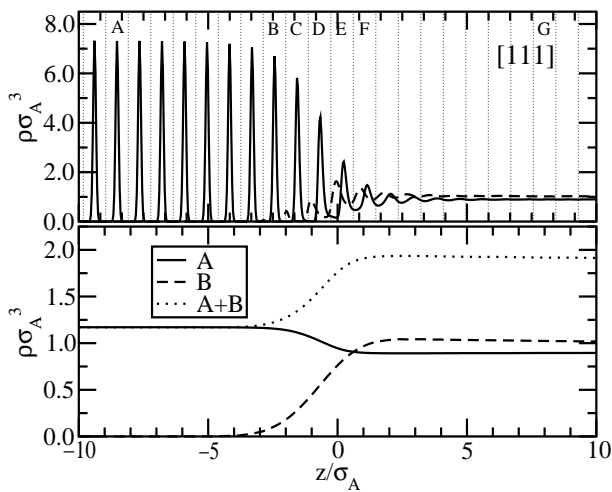}
\caption{\small Fine-scale (upper panel) and filtered (lower panel) density profiles 
for the [111] orientation.  The solid line and dashed lines are for the larger (A) and 
smaller (B) particles, respectively.  In the lower panel the dotted line shows the 
filtered total density. \hfill}
\label{f111rho}
\end{figure}

The fine scale density profiles for the [100] and [111] interfaces are 
shown in the upper panels of Figures~\ref{f100rho} and \ref{f111rho}, respectively. 
Shown in the lower panels are the corresponding filtered profiles (including the
total density profile). The distance along the $z$-axis (in units of the large
particle diameter, $\sigma_A$) is measured relative to the interface center, defined as discussed 
above by the orientational order profiles. The vertical dotted lines are equally spaced and constructed
to correspond with the density minima in between the bulk crystal layers. 
In both figures, specific interfacial layers are labeled alphabetically  for later reference - layers A and 
G correspond to bulk crystal and liquid, respectively, and layers B-F lie within the interfacial region.

The density profiles for the large particles resemble strongly those for the single component
hard sphere interface\cite{Davidchack98} with the periodic oscillations of the bulk crystal transforming
to the uniform density of the fluid over about 7-9 lattice layers as the interface is
traversed along the $z$-axis. The new feature seen in the present simulation is the decay of the small particle
density over a similar distance into the bulk crystal, in which the small particle are immiscible. As the
small particle density decreases into the crystal, it develops oscillations with a wavelength closely
matching that of the crystal lattice spacing.  For the [100] interface, the oscillations in the 
small particle density, $\rho_B(z)$ line up in phase with those of the large particle density,$\rho_A(z)$; whereas,
in the [111] interface the oscillations are out of phase - the peaks of $\rho_B(z)$ correspond to
minima of $\rho_A(z)$. Analysis of the atomic positions indicate that this difference is due to the
fact that in the interfacial region the small particles occupy interstitial sites of the large particle
fcc lattice - corresponding to the positions that would be occupied in an NaCl structure. These preferred
positions lie in the [100] plane, but lie between the [111] planes of the bulk fcc lattice. Recall that the
NaCl structure is the stable structure for this system at high pressure, so this effect is reminiscent of
premelting transitions at  solid/vapor interfaces below the bulk melting point, in that the presence
of a nearby triple point (in this case the fcc/NaCl/fluid triple point) manifests itself in the presence
of the metastable phase (NaCl) at the interface between the two 
coexisting phases (fcc and fluid). 

\begin{figure}[h]
\epsfig{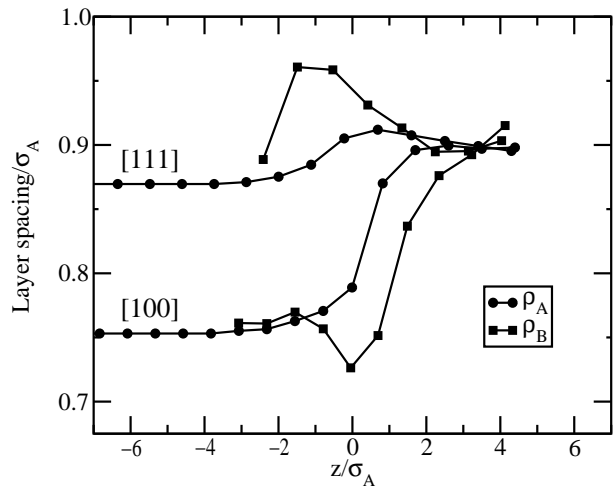}
\caption{Peak spacing as determined from maxima of fine-scale 
density profiles for both interfaces studied. \hfill}
\label{fspc}
\end{figure}

As in the single component hard-sphere system\cite{Davidchack98}, 
the spacings between the density peaks exhibit some 
variation across the interface - especially for the [100] orientation.
For each interface, the peak spacing was measured  by determining the distance between density peaks in the 
fine scale profiles.  The resulting peak spacings as functions of $z$  are shown in 
Fig.~\ref{fspc}. For the large particles the dependence of the spacing on interfacial orientation and
$z$ is identical to that seen in the single component simulations\cite{Davidchack98}. The spacing 
for the [100] lattice increases by nearly 20\% from the bulk crystal value of 0.76$\sigma_A$ to the
limiting value of about 0.9$\sigma_A$ as the bulk fluid is approached. The spacing for the large
particles in the [111] interface has the same bulk liquid limiting value, but since the bulk crystal
spacing is  very close to this limiting value, the variation in spacing across the interface is 
quite small.  The changes in peak spacing for the small particles are quite different for the 
different orientations and loosely follow those of the large particle - in [100] the small and
large particle curves have very similar shape, but are shifted by about $\sigma_A$. 

\begin{figure}[t]
\epsfig{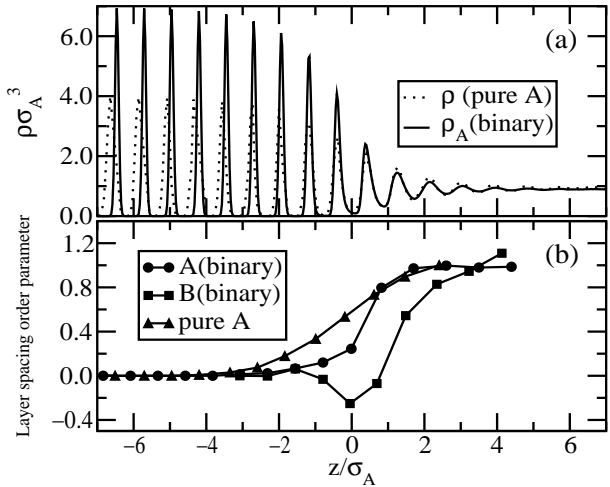}
\caption{\small Comparison of the binary interface with the previously 
studied hard-sphere single component simulation\cite{Davidchack98}. 
The upper panel shows the [100] fine scale density for both interfaces. The single
component data was shifted along the $z$-axis slightly to maximize the peak overlap in the 
fluid phase. The lower panel shows a comparison with the lattice spacing of the [100] interface -
for comparison purposes, the data is scaled and shifted (vertically) so that all curves go from 
zero in the crystal to unity in the fluid.)}
\label{fspc100}
\end{figure}

It is useful to compare these results directly with the single component case\cite{Davidchack98}. 
In Fig.~\ref{fspc100} we plot (upper panel) the fine scale density profiles for the [100] 
orientation of both the single component and binary interfaces. The single component data
was shifted slightly along $z$ to make the liquid peaks commensurate.  From this plot one
sees that the presence of the small particles has negligble effect on the coexisting liquid
density and structure; however, the higher pressure for the binary coexistence does give a 
crystal phase with a higher density (the peaks are more closely spaced and more localized).
The close similarity to the single component system indicates that the structure for the large
particles is changed very little  due to the presence of the smaller ones - except for
the higher density of the crystal. In the lower panel of Fig.~\ref{fspc100} is shown the
peak spacing for the [100] single component and binary  interfaces - scaled and shifted
so that the curves go from zero in the crystal to unity in the fluid. The curves 
for the large particles are qualitatively similar, but the change in the single component
case is less abrupt than that of the binary system.

A convenient measure of the width of the interfacial region is the so-called 10-90 width defined as the 
distance over which  an interfacial profile changes from  $10\%$ to $90\%$ 
of the higher of the two coexisting  bulk values relative to the lower bulk value.   Such a definition
is only useful for those interfacial profiles which are monotonic across the interface, such as
a coarse-grained (filtered) density or diffusion constants.  For the filtered large particle densities 
the 10-90 widths are $2.6\sigma_A$ for the [100] and $2.4 \sigma_A$ for the 
[111] - these are lower by about 0.8$\sigma_A$ than those found for the single component 
system\cite{Davidchack98} which were about 3.3$\sigma$ for the two interfaces.  From the small particle 
densities, the widths are larger at $3.4 \sigma_A$ and $3.2 \sigma_A$ for the [100] and [111] 
orientations, respectively, The 10-90 region defined by the large particles is within that defined 
by the small particles. The larger 10-90 width of the small particle filtered density is due to
the ability of the small particles to penetrate into the first few crystal lattice layers.

\begin{figure}[!h]
\epsfig{file=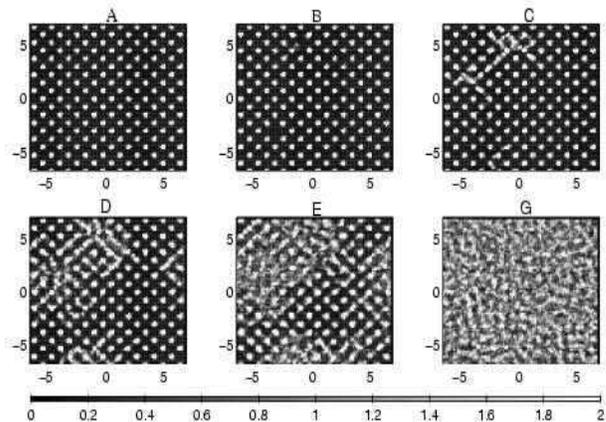,height=6.182cm,width=8.0cm}
\caption{Cross-sectional ({\it x-y}) density distributions of the large spheres for 
different layers of the [100] interface.}
\label{f100xyrha}
\end{figure}

\begin{figure}[h]
\epsfig{file=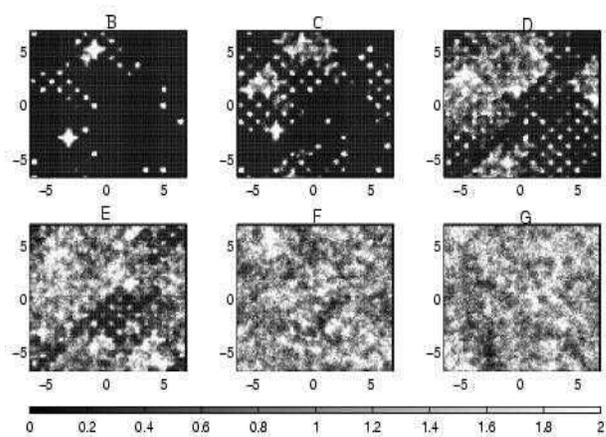,height=6.182cm,width=8.0cm}
\caption{Cross-sectional ({\it x-y}) density distributions of the small spheres for 
different layers of the [100] interface.}
\label{f100xyrhb}
\end{figure}

To get a more detailed picture of the transition from crystal-like to fluid-like structure as
the interface is traversed it is useful to examine the density distributions within {\it x-y} 
cross-sectional planes parallel to the interface.  (The reported distributions are 
averages taken over 1800 {\it cpp} -  details of their calculation can be found in the previous section.)  
Figures~\ref{f100xyrha} and ~\ref{f100xyrhb} respectively 
show the {\it x-y} large and small particle density distributions 
for the [100] interface orientation as greyscale contour plots.
The layer labels A-G  correspond to those shown in Fig.~\ref{f100rho}.  
Fig.~\ref{f100xyrha} shows  that this transition from crystal to fluid occurs over about three layers (C,D and E) 
for the [100] interface and that these transition layers are not uniform, but consist of coexisting
solid- and liquid-like regions, as was seen in the single-component simulations\cite{Davidchack98}. Layer B,
although fully crystalline, does possess two vacancy defects at points $(-3.3,-3.0)$ and $(-1.0,5.3)$.
The [100] contour plots for the small particle density are quite interesting. There is considerable 
density in layer B where the small particles are present in two types of positions - in the 
'NaCl' interstitial positions and in the positions corresponding to the vacancies of the 
large particle crystal lattice found in layer B. The interstitial positions are occupied by single
small particles, but each vacancy is filled with several small particles. 
In the single component simulations\cite{Davidchack98} vacancy nucleation at the interface was also 
seen, in that case the vacancies once formed were highly mobile, migrating into the bulk via a hopping mechanism.
In the present simulations, however, once the vacancies are formed in the large particle lattice, they 
are quickly filled with some number of small particles, which appears to 
immobilize the defect by suppressing the hopping mechanism - however the evidence for this
is anecdotal, as the number of such vacancies is too small to gather 
meaningful statistics. 

To estimate the degree of interfacial segregation, the Gibbs dividing surface for both
interfacial orientations was determined according to Eq.~\ref{gibbs} and found it to be close to 
the interface location determined from the orientational order parameters.  
The surface is at $z=-0.5\sigma_A$ for [100] and at $z=-0.9\sigma_A$ 
for [111].  At these dividing surfaces, the excess density of solute (here defined as component B) was
found to be negligible - indicating minimal  interfacial segregation. Of course, for such interfacial 
simulations, the question of complete chemical equilibrium is generally problematic, as discussed in the
previous section; however, we are confident that the concentrations of each particle type from interfacial
layer B out to the bulk fluid are in chemical equilibrium (since diffusion is non-negligible there) and
that the equilibrium concentrations of small particles in layers deeper into the crystal are probably
quite small and will not significantly affect the results presented here.

\subsection{Dynamics}

\begin{figure}[!h]
\epsfig{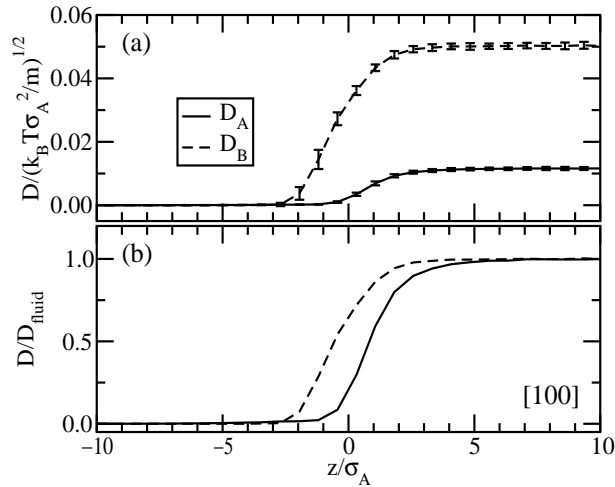}
\caption{(a) Diffusion coefficient profile for the [100] 
interface. (b) Scaled diffusion coefficients. \hfill}
\label{f100dif}
\end{figure}

\begin{figure}[h]
\epsfig{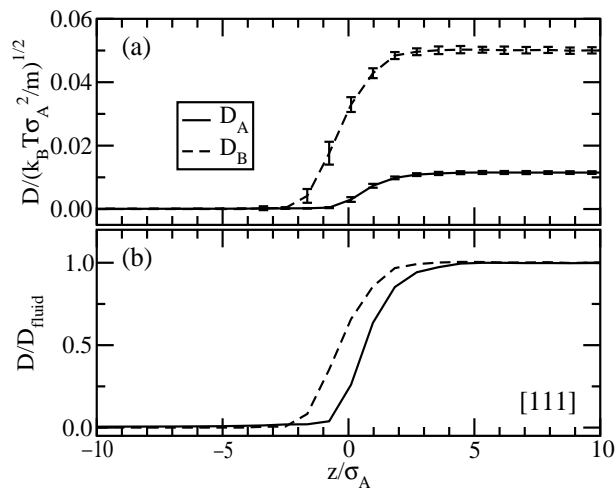}
\caption{(a) Diffusion coefficient profile for the [111] 
interface. (b) Scaled diffusion coefficients. \hfill}
\label{f111dif}
\end{figure}

We study the dynamics across the interface by measuring diffusion 
coefficients in the coarse-scaled bins.  The diffusion profiles for the [100] and
[111] interfaces are shown in Figures~\ref{f100dif}(a) and ~\ref{f111dif}(a), respectively.

The limiting bulk diffusion coefficient is $0.012 (k_BT\sigma_A^2/m)^{1/2}$ for the 
large spheres and  $0.050 (k_BT\sigma_A^2/m)^{1/2}$ for the small particles, independent 
of the crystal orientation, as expected.   When the three Cartesian components of the 
total diffusion coefficient are separately determined, it is found that 
diffusion is isotropic throughout the interfacial region. 

The larger value of the small
particle diffusion constant makes it difficult to compare the diffusion constants of the
two components so we also plot for each, the ratio diffusion constant to the average fluid bulk 
value in Figures~\ref{f100dif}(b) and ~\ref{f111dif}(b). Here we find the
interesting result that the two curves (for both crystal orientations) are similar in
shape, but shifted relative to one another by more than 1$\sigma_A$. As the interface is traversed 
from fluid to crystal, the diffusion constant for the large particle goes effectively to zero near $z=0$, but
the small particles still have significant mobility.  In this region, the large particles have become
``locked in'' to their crystal lattice sites, but the small particles can still move about - primarily by
hopping between interstitial sites.

\begin{figure}[h]
\epsfig{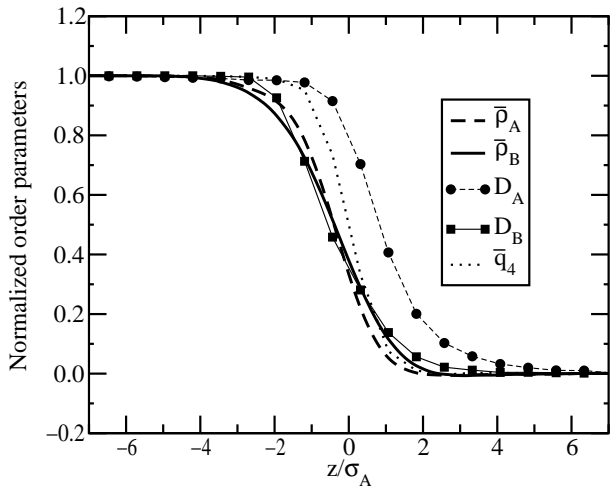}
\caption{\small Diffusion, orientation and filtered density 
order parameter profiles for the [100] interface - all scaled such that they go from
unity in the crystal to zero in the fluid phase. \hfill}
\label{f100wid}
\end{figure}

The 10-90 widths from the diffusion coefficient profiles for both 
orientation and particle types are about $3 \sigma_A$.    
But because the diffusion profiles are shifted, the 10-90 widths do not 
define the same region.  If contributions from both particle types 
are considered, the widths are $4.5 \sigma_A$ for the [100] and 
$3.9 \sigma_A$ for the [111] interface.  The center of these interfacial 
regions are shifted by about $1 \sigma_A$ to the fluid side compared to the 
interfacial regions defined by the density profiles. 
To illustrate this more clearly, we show in Fig.~\ref{f100wid} all of the 
order parameter profiles (orientation, diffusion and density) for the [100]
interface, scaled in such a way that they go from unity in the crystal phase to zero in 
the liquid (for example for the diffusion constants we plot $1 - D(z)/D_f$). 
 
The 10-90 regions for the diffusion constants are offset (toward the liquid side) from those for the filtered density
profiles so the interfacial region is wider than any single structural or dynamical quantity would 
indicate. If one considers the interfacial region as the union of the 10-90 regions for the separate profiles,
then the width of the interfacial region is $4.8 \sigma_A$, greater than that calculated from 
densities or diffusion coefficients alone.

\section{Summary}

We have performed a series of molecular-dynamics simulations to study the crystal-melt
interface of a binary hard-sphere system with diameter ratio 0.414. Previous simulation
studies on two-component crystal-melt interfaces have focused on equilibrium between
a fluid mixture and a substitutionally disordered crystal\cite{Davidchack96,Davidchack99}, 
but here we have examined the interface between a fluid mixture (approximately equimolar in
concentration) and a coexisting {\it single}-component fcc crystal comprised of large particles - in which the
small particles are immiscible.  Such a coexistence occurs at relatively low pressures 
in the phase diagram for this diameter ratio - at higher pressures the fluid coexists
with a 1:1 ordered crystal with an `NaCl' structure. At a pressure of
$P=20.1 \sigma_A^3/kT$ the two phases coexist at the following packing fractions:  $\eta_c=0.61$ and $\eta_f=0.51$.

Some of the  principal results of this study are as follows:
\begin{itemize}
\item The interfacial density profiles of the large particles is very similar to
that of the single-component hard-sphere system previously studied\cite{Davidchack98}, indicating
that the presence of the small particle has no significant effect on the interfacial structure
of the large particle, except for a compression of the crystal lattice due to the higher pressure.  
In particular the variation of the spacing between the large particle
density peaks is very similar to that found in the single component studies.
\item Within the regions of the interface in which the large particles are largely confined
to fcc lattice sites, the small particles occupy either vacancy sites in the fcc lattice 
or `NaCl' interstitial sites. The interstitial sites are single occupied whereas the vacancy sites
are found to be occupied by several small particles. The presence of the small particles greatly
suppresses the mobility of the fcc vacancies relative to those previously noted in single-component 
hard-sphere  simulations\cite{Davidchack98}.
\item There does not appear to be significant solute (small particle)  segregation at the
interface. 
\item The diffusion profiles of the small and large particles are similar in width (about
3 $\sigma_A$), but are shifted relative to one another by about 1 $\sigma_A$ along the
interface normal ($z$ axis). As one traverses the interface from bulk fluid to bulk crystal,
the diffusion constant goes to zero for the large particles in a region in which there
is still significant small particle mobility. The picture in this region is of large particles
localized at fcc lattice sites, with the small particles still diffusing between interstitial 
site within the lattice of large particles. 
\item  As was found in previous hard-sphere interface studies\cite{Davidchack98,Davidchack99}
the total width of the interfacial region is greater than the width determined by any single
interfacial profile (such as diffusion or density) as the profiles for the individual quantities
can be significantly shifted from one another. Specifically we see that as one moves from
the crystal into the fluid the bulk density relaxes first to liquid-like values before
significant mobility (diffusion) is observed.  Considering both  structural and dynamic 
properties, the interfacial (10-90) width is $4.8\sigma_A$.
\end{itemize}

\section{Acknowledgements} \label{sec:Acknowledgements}
We gratefully acknowledge R.L. Davidchack for helpful conversations,
as well as the Kansas Center for Advanced Scientific Computing for
the use of their computer facilities. We also would like to thank the
National Science Foundation for generous support under grant CHE-9500211. 

\newpage


\begin{thebibliography}{25}
\expandafter\ifx\csname natexlab\endcsname\relax\def\natexlab#1{#1}\fi
\expandafter\ifx\csname bibnamefont\endcsname\relax
  \def\bibnamefont#1{#1}\fi
\expandafter\ifx\csname bibfnamefont\endcsname\relax
  \def\bibfnamefont#1{#1}\fi
\expandafter\ifx\csname citenamefont\endcsname\relax
  \def\citenamefont#1{#1}\fi
\expandafter\ifx\csname url\endcsname\relax
  \def\url#1{\texttt{#1}}\fi
\expandafter\ifx\csname urlprefix\endcsname\relax\def\urlprefix{URL }\fi
\providecommand{\bibinfo}[2]{#2}
\providecommand{\eprint}[2][]{\url{#2}}

\bibitem{brian} Author to whom correspondence should be addressed. 
\bibitem[{\citenamefont{Woodruff}(1973)}]{Woodruff73}
\bibinfo{author}{\bibfnamefont{D.}~\bibnamefont{Woodruff}},
  \emph{\bibinfo{title}{The Solid-Liquid Interface}}
  (\bibinfo{publisher}{Cambridge University Press}, \bibinfo{address}{London},
  \bibinfo{year}{1973}).

\bibitem[{\citenamefont{Tiller}(1991)}]{Tiller91}
\bibinfo{author}{\bibfnamefont{W.}~\bibnamefont{Tiller}},
  \emph{\bibinfo{title}{The Science of Crystallization: Microscopic Interfacial
  Phenomena}} (\bibinfo{publisher}{Cambridge University Press},
  \bibinfo{address}{New York}, \bibinfo{year}{1991}).

\bibitem[{\citenamefont{Howe}(1997)}]{Howe97}
\bibinfo{author}{\bibfnamefont{J.}~\bibnamefont{Howe}},
  \emph{\bibinfo{title}{Interfaces in Materials}} (\bibinfo{publisher}{John
  Wiley \& Sons}, \bibinfo{address}{New York}, \bibinfo{year}{1997}).

\bibitem[{\citenamefont{Adamson and Gast}(1997)}]{Adamson97}
\bibinfo{author}{\bibfnamefont{A.}~\bibnamefont{Adamson}} \bibnamefont{and}
  \bibinfo{author}{\bibfnamefont{A.}~\bibnamefont{Gast}},
  \emph{\bibinfo{title}{Physical Chemistry of Surfaces}}
  (\bibinfo{publisher}{Wiley-Interscience}, \bibinfo{address}{New York},
  \bibinfo{year}{1997}).

\bibitem[{\citenamefont{Laird}(1998)}]{Laird98}
\bibinfo{author}{\bibfnamefont{B.}~\bibnamefont{Laird}}, in
  \emph{\bibinfo{booktitle}{Encyclopedia of Computational Chemistry}}, edited by
  \bibinfo{editor}{\bibfnamefont{P.}~\bibnamefont{Schleyer}},
  \bibinfo{editor}{\bibfnamefont{N.}~\bibnamefont{Allinger}},
  \bibinfo{editor}{\bibfnamefont{T.}~\bibnamefont{Clark}},
  \bibinfo{editor}{\bibfnamefont{P.}~\bibnamefont{Kollman}}, \bibnamefont{and}
  \bibinfo{editor}{\bibfnamefont{H.}~\bibnamefont{Schaefer}}
  (\bibinfo{publisher}{J. Wiley and Sons}, \bibinfo{address}{New York},
  \bibinfo{year}{1998}).

\bibitem[{\citenamefont{Kyrlidis and Brown}(1995)}]{Kyrlidis95}
\bibinfo{author}{\bibfnamefont{A.}~\bibnamefont{Kyrlidis}} \bibnamefont{and}
  \bibinfo{author}{\bibfnamefont{R.}~\bibnamefont{Brown}},
  \bibinfo{journal}{Phys. Rev. E} \textbf{\bibinfo{volume}{51}},
  \bibinfo{pages}{5832} (\bibinfo{year}{1995}).

\bibitem[{\citenamefont{Mori et~al.}(1995)\citenamefont{Mori, Manabe, and
  Nishioka}}]{Mori95}
\bibinfo{author}{\bibfnamefont{A.}~\bibnamefont{Mori}},
  \bibinfo{author}{\bibfnamefont{R.}~\bibnamefont{Manabe}}, \bibnamefont{and}
  \bibinfo{author}{\bibfnamefont{K.}~\bibnamefont{Nishioka}},
  \bibinfo{journal}{Phys. Rev. E} \textbf{\bibinfo{volume}{51}},
  \bibinfo{pages}{R3831} (\bibinfo{year}{1995}).

\bibitem[{\citenamefont{Davidchack and Laird}(2000)}]{Davidchack00}
\bibinfo{author}{\bibfnamefont{R.}~\bibnamefont{Davidchack}} \bibnamefont{and}
  \bibinfo{author}{\bibfnamefont{B.}~\bibnamefont{Laird}},
  \bibinfo{journal}{Phys. Rev. Lett.} \textbf{\bibinfo{volume}{85}},
  \bibinfo{pages}{4751} (\bibinfo{year}{2000}).

\bibitem[{\citenamefont{Broughton and Gilmer}(1986)}]{Broughton86c}
\bibinfo{author}{\bibfnamefont{J.}~\bibnamefont{Broughton}} \bibnamefont{and}
  \bibinfo{author}{\bibfnamefont{G.}~\bibnamefont{Gilmer}},
  \bibinfo{journal}{J. Chem. Phys.} \textbf{\bibinfo{volume}{84}},
  \bibinfo{pages}{5759} (\bibinfo{year}{1986}).

\bibitem[{\citenamefont{Galejs et~al.}(1989)\citenamefont{Galejs, Raveche, and
  Lie}}]{Galejs89}
\bibinfo{author}{\bibfnamefont{R.}~\bibnamefont{Galejs}},
  \bibinfo{author}{\bibfnamefont{H.}~\bibnamefont{Raveche}}, \bibnamefont{and}
  \bibinfo{author}{\bibfnamefont{G.}~\bibnamefont{Lie}},
  \bibinfo{journal}{Phys. Rev. A} \textbf{\bibinfo{volume}{39}},
  \bibinfo{pages}{2574} (\bibinfo{year}{1989}).

\bibitem[{\citenamefont{Karim and Haymet}(1988)}]{Karim88}
\bibinfo{author}{\bibfnamefont{O.}~\bibnamefont{Karim}} \bibnamefont{and}
  \bibinfo{author}{\bibfnamefont{A.}~\bibnamefont{Haymet}},
  \bibinfo{journal}{J. Chem. Phys.} \textbf{\bibinfo{volume}{89}},
  \bibinfo{pages}{6889} (\bibinfo{year}{1988}).

\bibitem[{\citenamefont{Karim et~al.}(1990)\citenamefont{Karim, Kay, and
  Haymet}}]{Karim90}
\bibinfo{author}{\bibfnamefont{O.}~\bibnamefont{Karim}},
  \bibinfo{author}{\bibfnamefont{P.}~\bibnamefont{Kay}}, \bibnamefont{and}
  \bibinfo{author}{\bibfnamefont{A.}~\bibnamefont{Haymet}},
  \bibinfo{journal}{J. Chem. Phys.} \textbf{\bibinfo{volume}{92}},
  \bibinfo{pages}{4634} (\bibinfo{year}{1990}).

\bibitem[{\citenamefont{Hayward and Haymet}(2001)}]{Hayward01}
\bibinfo{author}{\bibfnamefont{J.}~\bibnamefont{Hayward}} \bibnamefont{and}
  \bibinfo{author}{\bibfnamefont{A.}~\bibnamefont{Haymet}},
  \bibinfo{journal}{J. Chem. Phys.} \textbf{\bibinfo{volume}{114}},
  \bibinfo{pages}{3713} (\bibinfo{year}{2001}).

\bibitem[{\citenamefont{Abraham and Broughton}(1986)}]{Abraham86}
\bibinfo{author}{\bibfnamefont{F.}~\bibnamefont{Abraham}} \bibnamefont{and}
  \bibinfo{author}{\bibfnamefont{J.}~\bibnamefont{Broughton}},
  \bibinfo{journal}{Phys. Rev. Lett.} \textbf{\bibinfo{volume}{56}},
  \bibinfo{pages}{734} (\bibinfo{year}{1986}).

\bibitem[{\citenamefont{Landman et~al.}(1986)\citenamefont{Landman, Luedtke,
  Barnett, Cleveland, Ribarsky, Arnold, Ramesh, Baumgart, Martinez, and
  Khan}}]{Landman86}
\bibinfo{author}{\bibfnamefont{U.}~\bibnamefont{Landman}},
  \bibinfo{author}{\bibfnamefont{W.}~\bibnamefont{Luedtke}},
  \bibinfo{author}{\bibfnamefont{R.}~\bibnamefont{Barnett}},
  \bibinfo{author}{\bibfnamefont{C.}~\bibnamefont{Cleveland}},
  \bibinfo{author}{\bibfnamefont{M.}~\bibnamefont{Ribarsky}},
  \bibinfo{author}{\bibfnamefont{E.}~\bibnamefont{Arnold}},
  \bibinfo{author}{\bibfnamefont{S.}~\bibnamefont{Ramesh}},
  \bibinfo{author}{\bibfnamefont{H.}~\bibnamefont{Baumgart}},
  \bibinfo{author}{\bibfnamefont{A.}~\bibnamefont{Martinez}}, \bibnamefont{and}
  \bibinfo{author}{\bibfnamefont{B.}~\bibnamefont{Khan}},
  \bibinfo{journal}{Phys. Rev. Lett.} \textbf{\bibinfo{volume}{56}},
  \bibinfo{pages}{155} (\bibinfo{year}{1986}).

\bibitem[{\citenamefont{Jesson and Madden}(2001)}]{Jesson01}
\bibinfo{author}{\bibfnamefont{B.}~\bibnamefont{Jesson}} \bibnamefont{and}
  \bibinfo{author}{\bibfnamefont{P.}~\bibnamefont{Madden}},
  \bibinfo{journal}{J. Chem. Phys.} \textbf{\bibinfo{volume}{113}},
  \bibinfo{pages}{5935} (\bibinfo{year}{2001}).

\bibitem[{\citenamefont{Hoyt et~al.}(2001)\citenamefont{Hoyt, Asta, and
  Karma}}]{Hoyt01}
\bibinfo{author}{\bibfnamefont{J.}~\bibnamefont{Hoyt}},
  \bibinfo{author}{\bibfnamefont{M.}~\bibnamefont{Asta}}, \bibnamefont{and}
  \bibinfo{author}{\bibfnamefont{A.}~\bibnamefont{Karma}},
  \bibinfo{journal}{Phys. Rev. Lett} \textbf{\bibinfo{volume}{86}},
  \bibinfo{pages}{5530} (\bibinfo{year}{2001}).

\bibitem[{\citenamefont{Davidchack and Laird}(1999)}]{Davidchack99}
\bibinfo{author}{\bibfnamefont{R.}~\bibnamefont{Davidchack}} \bibnamefont{and}
  \bibinfo{author}{\bibfnamefont{B.}~\bibnamefont{Laird}},
  \bibinfo{journal}{Mol. Phys.} \textbf{\bibinfo{volume}{97}},
  \bibinfo{pages}{833} (\bibinfo{year}{1999}).

\bibitem[{\citenamefont{Davidchack and Laird}(1996)}]{Davidchack96}
\bibinfo{author}{\bibfnamefont{R.}~\bibnamefont{Davidchack}} \bibnamefont{and}
  \bibinfo{author}{\bibfnamefont{B.}~\bibnamefont{Laird}},
  \bibinfo{journal}{Phys. Rev. E} \textbf{\bibinfo{volume}{54}},
  \bibinfo{pages}{R5905} (\bibinfo{year}{1996}).


\bibitem[{\citenamefont{Laird}(2001)}]{Laird01}
\bibinfo{author}{\bibfnamefont{B.}~\bibnamefont{Laird}}, \bibinfo{journal}{J.
  Chem. Phys.} \textbf{\bibinfo{volume}{115}}, \bibinfo{pages}{2889}
  (\bibinfo{year}{2001}).

\bibitem[{\citenamefont{Trizac et~al.}(1997)\citenamefont{Trizac, Eldridge, and
  Madden}}]{Trizac97}
\bibinfo{author}{\bibfnamefont{E.}~\bibnamefont{Trizac}},
  \bibinfo{author}{\bibfnamefont{M.~D.} \bibnamefont{Eldridge}},
  \bibnamefont{and} \bibinfo{author}{\bibfnamefont{P.~A.}
  \bibnamefont{Madden}}, \bibinfo{journal}{Mol. Phys.}
  \textbf{\bibinfo{volume}{90}}, \bibinfo{pages}{675} (\bibinfo{year}{1997}).

\bibitem[{\citenamefont{Cottin and Monson}(1995)}]{Cottin95}
\bibinfo{author}{\bibfnamefont{X.}~\bibnamefont{Cottin}} \bibnamefont{and}
  \bibinfo{author}{\bibfnamefont{P.~A.} \bibnamefont{Monson}},
  \bibinfo{journal}{J. Chem. Phys.} \textbf{\bibinfo{volume}{102}},
  \bibinfo{pages}{3354} (\bibinfo{year}{1995}).

\bibitem[{\citenamefont{Davidchack and Laird}(1998)}]{Davidchack98}
\bibinfo{author}{\bibfnamefont{R.}~\bibnamefont{Davidchack}} \bibnamefont{and}
  \bibinfo{author}{\bibfnamefont{B.}~\bibnamefont{Laird}}, \bibinfo{journal}{J.
  Chem. Phys.} \textbf{\bibinfo{volume}{108}}, \bibinfo{pages}{9452}
  (\bibinfo{year}{1998}).

\bibitem[{\citenamefont{Press et~al.}(1992)\citenamefont{Press, Teukolsky,
  Vetterling, and Flannery}}]{NumRec}
\bibinfo{author}{\bibfnamefont{W.}~\bibnamefont{Press}},
  \bibinfo{author}{\bibfnamefont{S.}~\bibnamefont{Teukolsky}},
  \bibinfo{author}{\bibfnamefont{W.}~\bibnamefont{Vetterling}},
  \bibnamefont{and} \bibinfo{author}{\bibfnamefont{B.}~\bibnamefont{Flannery}},
  \emph{\bibinfo{title}{Numerical Recipies in Fortran}}
  (\bibinfo{publisher}{Cambridge University Press}, \bibinfo{address}{New
  York}, \bibinfo{year}{1992}).

\bibitem[{\citenamefont{Rappaport}(1995)}]{Rappaport95}
\bibinfo{author}{\bibfnamefont{D.~C.} \bibnamefont{Rappaport}},
  \emph{\bibinfo{title}{The Art of Molecular Dynamics Simulation}}
  (\bibinfo{publisher}{Cambridge University Press}, \bibinfo{address}{New
  York}, \bibinfo{year}{1995}).

\bibitem[{\citenamefont{Frenkel}(1996)}]{Frenkel96}
\bibinfo{author}{\bibfnamefont{D.}~\bibnamefont{Frenkel}}
  \bibnamefont{and} \bibinfo{author}{\bibfnamefont{B.}~\bibnamefont{Smit}},
  \emph{\bibinfo{title}{Understanding Molecular Simulation}}
  (\bibinfo{publisher}{Academic Press}, \bibinfo{address}{New York},
  \bibinfo{year}{1996}).

\end{thebibliography}
\end{document}